\documentclass[a4paper,11pt]{article}
\usepackage{pos,graphicx}

\def\simge{\mathrel{%
       \rlap{\raise 0.511ex \hbox{$>$}}{\lower 0.511ex \hbox{$\sim$}}}}
\def\simle{\mathrel{
       \rlap{\raise 0.511ex \hbox{$<$}}{\lower 0.511ex \hbox{$\sim$}}}}

\title{Phase structure and critical point in heavy-quark QCD at finite temperature%
 \\ \vspace*{-62mm}\hfill \small{\texttt{UTHEP-775, YITP-22-131, J-PARC-TH-0280, KYUSHU-HET-251}} \vspace*{58mm}
}
\ShortTitle{Phase structure and critical point in heavy-quark QCD} 

\author*[a]{Kazuyuki Kanaya}
\author[b]{Ryo Ashikawa}
\author[c]{Shinji Ejiri}
\author[d,e,b]{Masakiyo Kitazawa}
\author[f]{\\Hiroshi Suzuki}
\author[g]{Naoki Wakabayashi}

\affiliation[a]{Tomonaga Center for the History of the Universe, University of Tsukuba, \\Tsukuba, Ibaraki 305-8571, Japan}

\affiliation[b]{Department of Physics, Osaka University, Toyonaka, Osaka 560-0043, Japan}

\affiliation[c]{Department of Physics, Niigata University, Niigata 950-2181, Japan}

\affiliation[d]{Yukawa Institute for Theoretical Physics, Kyoto University, Kyoto 606-8502, Japan}

\affiliation[e]{J-PARC Branch, KEK Theory Center, Institute of Particle and Nuclear Studies, KEK, \\203-1 Shirakata, Tokai, Ibaraki, 319-1106, Japan}

\affiliation[f]{Department of Physics, Kyushu University, 744 Motooka, Nishi-ku, Fukuoka 819-0395, Japan}

\affiliation[g]{Graduate School of Science and Technology, Niigata University, Niigata 950-2181, Japan}

\emailAdd{kanaya@ccs.tsukuba.ac.jp}

\abstract{We study phase structure and critical point of finite-temperature QCD in the heavy-quark region applying the hopping parameter expansion (HPE). 
We first study finite-size scaling on the critical point on $N_t=4$ lattices with large spatial volumes taking the leading order (LO) and the next-to-leading order (NLO) effects of the HPE, and find that the critical scaling of the Z(2) universality class expected around the critical point of two-flavor QCD is realized when the aspect ratio of the lattice is larger than about 9. 
This enables us to determine the critical point in the thermodynamic limit with high precisions. By a study of the convergence of the HPE, we confirm that the result of the critical point with the LO (NLO) approximation of the HPE is fairly accurate for $N_t=4$ (6), while we need to incorporate higher order effects for larger $N_t$. 
To extend the study to large $N_t$ lattices, we then develop a method to take the effects of higher-order terms of the HPE up to a sufficiently high order. 
We report on the status of our study on $N_t = 6$ lattice adopting the new method.
}

\FullConference{%
The 39th International Symposium on Lattice Field Theory,\\
8th-13th August, 2022,\\
Rheinische Friedrich-Wilhelms-Universit\"at Bonn, Bonn, Germany
}

\begin{document}
\maketitle

\section{Critical quark mass in heavy-quark QCD}

Around 0.1 milli-second after the Big Bang, our Universe has experienced a phase transition from the hot quark matter to ordinary matters such as protons and neutrons.
This QCD transition plays a decisive role in the initial condition for the generation and evolution of ordinary matters in our Universe.
The QCD transition is known to be an analytic crossover when all quark masses are set to their physical values.
Though other points in the parameter space of quark masses are unphysical, 
clarification of the phase structure of QCD off the physical point is also important as the properties at the physical point may be affected by the scaling of nearby critical points.

When all quarks are infinitely heavy, QCD tends to the SU(3) pure gauge Yang-Mills theory, 
which has a first-order phase transition between the low-temperature confined phase and the high-temperature deconfined phase. 
The transition can be effectively described by the Z(3) Potts model, in which the Z(3) spin variables correspond to the Polyakov loops.
When we decrease the quark masses from infinity, because quarks act as external fields to the Z(3) spin system, this first-order transition weakens and eventually turns into a crossover at some critical quark mass.
In this paper, we report on our recent studies of the critical quark mass in QCD with heavy quarks \cite{Kiyohara:2021smr,wakabayashi,ashikawa}.
While the first-order nature of the heavy quark limit is well established, determination of the critical quark mass is still a delicate issue.
Recent lattice studies on the location of the critical quark mass showed that we still have strong cutoff and spatial volume dependences in the result \cite{Saito:2011fs,Saito:2013vja,Ejiri:2019csa,Cuteri:2020yke,Kara:2021btt}.

We first have to remove the spatial volume dependence.
A powerful way to determine the critical point in the infinite volume limit is to study the Binder cumulant, which is designed to remove the leading volume-dependence of the finite-size scaling (FSS) when we tune the parameter to the critical point. 
Thus the point where the Binder cumulant measured with sufficiently large spatial volumes becomes insensitive to the volume gives a good estimate of the critical point in the infinite volume limit. 
In ref.~\cite{Cuteri:2020yke}, the Binder cumulant of the Polyakov loop was studied on $N_t=6$, 8, and 10 lattices with the spatial volumes corresponding to the aspect ratio $N_s/N_t=4$--7 (10).
It was reported that the high-temperature data deviate from leading FSS fits, and, even after removing them, the crossing point of the Binder cumulant moves as we increase the spatial volume. 
These procedures make the scaling analyses delicate and call for a careful estimation of systematic errors. 
To identify the leading FSS more clearly, we thus decided to carry out simulations with larger spatial volumes.
To achieve high statistics with large spatial volumes, we adopt the hopping parameter expansion (HPE) in this study. 
We also adopt the multi-point reweighting method to vary coupling parameters continuously in the Binder cumulant study.
%

\section{Lattice setup and HPE}
\label{sec:setup}

Simulations are performed on $N_s^3\times N_t$ lattice with the lattice spacing $a$. 
For each temperature $T=1/(N_t a)$, we measure the spatial lattice size $L=N_sa$ in terms of the aspect ratio $N_s/N_t=LT$.
We adopt the standard plaquette gauge action and the standard Wilson quark action. 
The quark kernel is given by 
\begin{equation}
  M_{xy} (\kappa_f) = \delta_{xy} - \kappa_f B_{xy},
\hspace{2mm} \textrm{with} \;\;
  B_{xy}
  =  \sum_{\mu=1}^4 \left[ (1-\gamma_{\mu})\,U_{x,\mu}\,\delta_{y,x+\hat{\mu}} + (1+\gamma_{\mu})\,U_{y,\mu}^{\dagger}\,\delta_{y,x-\hat{\mu}} \right],
  \label{eq:B}
\end{equation} 
where $x$, $y$ represent lattice sites and $\kappa_f = 1/(2am_f+8)$ is the hopping parameter for the $f$th flavor with the bare quark mass $m_f$. 
Then the quark contribution to the effective action from the $f$th flavor can be expanded as
\begin{equation}
\ln \det M(\kappa_f) = N_{\textrm{site}} \sum_{n=1}^{\infty} D_n \kappa_f^n,
\hspace{5mm}
D_n =  \frac{-1}{N_{\textrm{site}}n} \textrm{Tr}\,[B^n] = W(n) + L(N_t,n) .
  \label{eq:HPE}
\end{equation} 
Here, non-vanishing contributions to $\textrm{Tr}\,[B^n]$ are given by closed loops of the hopping term $B$ with the length $n$, which are classified to contributions of Wilson-type loops $W(n)$ and those of Polyakov-type loops $L(N_t,n)$. 
The latter can be further decomposed as $L(N_t,n) = \sum_m L_m(N_t,n)$ where $m$ is the winding number in the temporal direction. 
In the following, we mainly consider the case of degenerate $N_f$ flavors, though generalization to non-degenerate cases is straightforward. 

To the leading order (LO) of the Wilson-type and the Polyakov-type contributions, we have  
\begin{equation}
W(4) \, \kappa^4 = 
288 \, \kappa^4\, \hat{P},
\hspace{5mm}
L(N_t,N_t) \, \kappa^{N_t} = 
\frac{12}{N_t} 2^{N_t}  \kappa^{N_t} 
 \textrm{Re}\,\hat{\Omega} ,
  \label{eq:LO}
\end{equation} 
respectively,
where $\hat{P}$ is the plaquette and $\hat{\Omega}$ is the Polyakov loop, both averaged over space-time position on each configuration and normalized such that they take their maximum value 1 when we set all link variables to unity\footnote{
We use the term ``LO'' in the sense of Eq.~(\ref{eq:LO}), i.e., the LO Polyakov-type contribution is $O(\kappa^{N_t})$.
}. 
Similarly, to the next-to-leading order (NLO), $W(6)$ is given in terms of length-6 Wilson-type loops and $L(N_t,N_t+2)$ is given in terms of bent Polyakov loops of length $N_t+2$. 
See \cite{Kiyohara:2021smr,wakabayashi} for details.

In this study, we incorporate the LO contributions in the configuration generation. 
Effects of $W(4)$ can be absorbed simply by shifting the gauge coupling parameter: 
$\beta \rightarrow \beta^* = \beta + 48 N_f \kappa^4$.
The Polyakov-loop contribution $L(N_t,N_t)$ induces the term 
\begin{equation}
N_s^3 \lambda \, \textrm{Re} \,\hat\Omega
\end{equation} 
in the effective action,
where $\lambda=48N_fN_t\kappa^{N_t}$ for $N_t=4$ and $128N_fN_t\kappa^{N_t}$ for $N_t=6$ \cite{wakabayashi}.
This term can be incorporated in the parallel pseudo-heat-bath simulation algorithm and thus can be simulated without increasing the simulation cost much from the quenched QCD simulation.
We then incorporate the NLO effects through the multi-point reweighting procedure in the measurements.
See \cite{Kiyohara:2021smr} for details.
Note that, to the LO approximation,  we can directly generate configurations at finite $\lambda$. 
We find that this helps much to remove the overlap problem in the reweighting method, which is quite severe on large lattices as study in this paper.

\section{Results at $N_t=4$ \cite{Kiyohara:2021smr}}
\label{sec:nt4}

\begin{figure}
  \centering
    \includegraphics[width=0.48\textwidth]{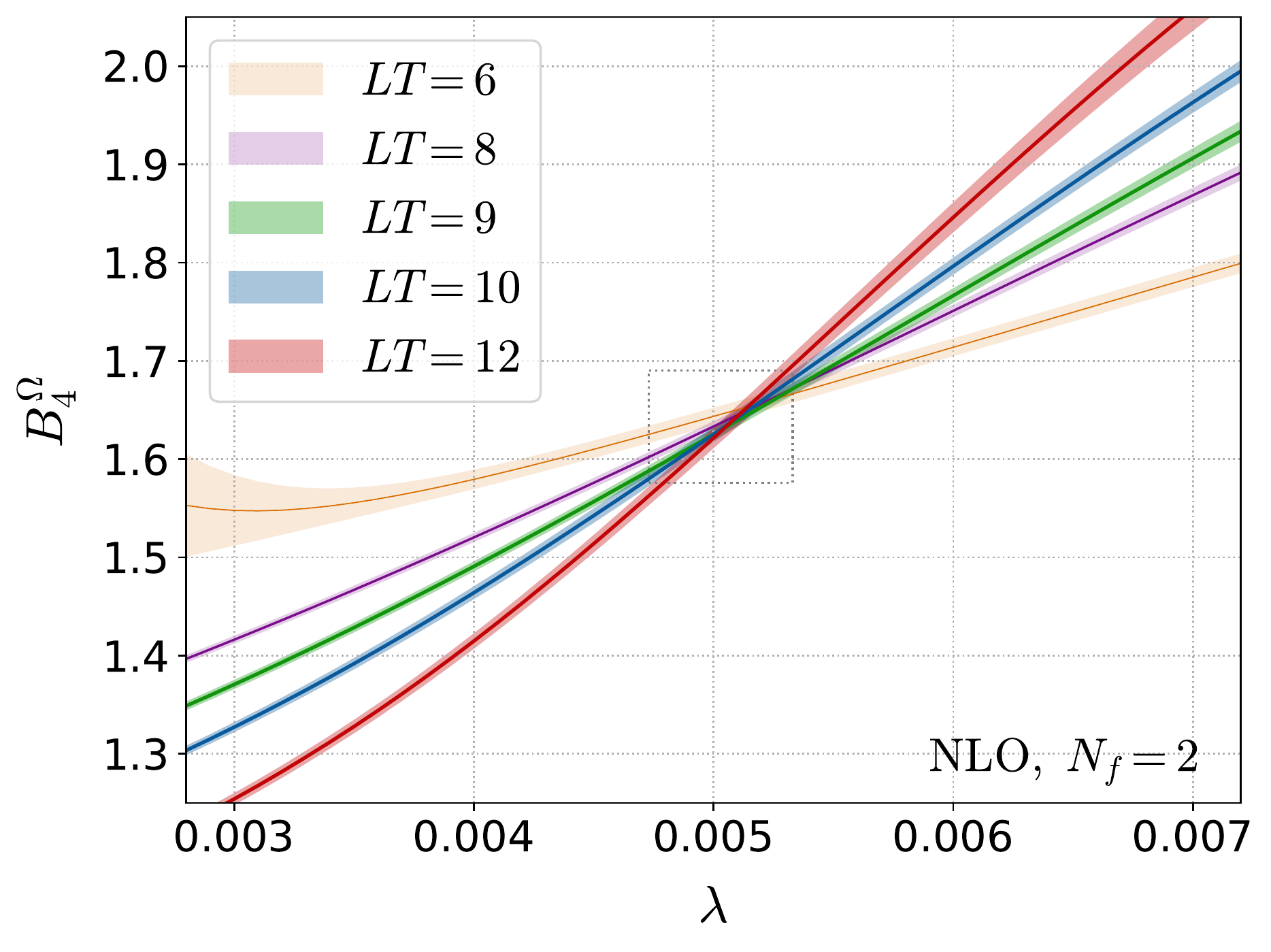}
    \hspace{1mm}
    \includegraphics[width=0.48\textwidth]{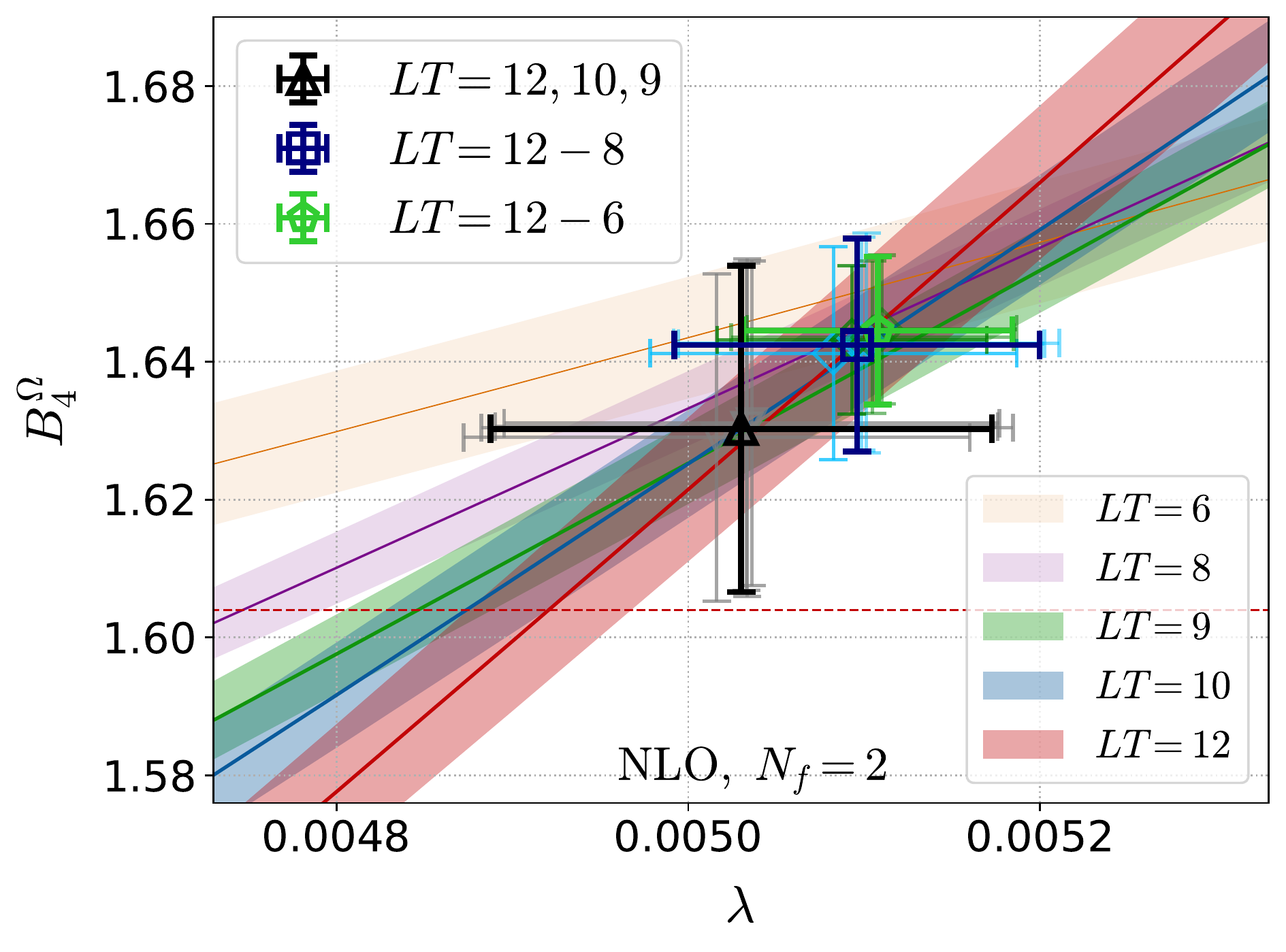}
    \vspace*{-0.5mm}
  \caption{
    Binder cumulant $B_4^\Omega$ on $N_t=4$ lattices at five aspect ratios $LT=N_s/N_t$ \cite{Kiyohara:2021smr}. 
    The statistical errors are shown by the shaded area.
    Terms up to the NLO of HPE are incorporated.
    The right panel is an enlargement of the left panel around the crossing point (the dotted rectangle region).
    The points in the right panel with error bars show the results of a FSS fit.
(Thin symbols with pale colors are for alternative choices of $\lambda$ in the fit.
See \cite{Kiyohara:2021smr} for fitting and error estimation details.)
  }
\label{fig:B4}
\end{figure}

Because simulation with large spatial volumes is costly, we first revisit the case of $N_t=4$  \cite{Kiyohara:2021smr}.
We perform simulations at several $\beta$'s and $\kappa$'s ($\lambda$'s) around the transition line 
on lattices with spatial lattice sizes $N_s=24$, 32, 36, 40, and 48, corresponding to the aspect ratio $N_s/N_t = LT= 6$--12. 
Using about $10^6$ configurations at each simulation point, we compute the Binder cumulant for the real part of the Polyakov loop $\hat\Omega$:
 \begin{equation}
  B_4^\Omega
  = \frac{ \langle \Omega_{\rm R}^4 \rangle_{\rm c} }
  { \langle \Omega_{\rm R}^2 \rangle_{\rm c}^2 } + 3 
  , \hspace{5mm}
\Omega_{\rm R} =
{\rm Re}\,\hat\Omega = \frac1{N_{\rm c}N_s^3}
 \sum_{\vec{x}} {\rm Re}\,{\rm tr_C} \left[ 
U_{\vec{x},4} U_{\vec{x}+\hat{4},4} U_{\vec{x}+2 \cdot \hat{4},4} 
\cdots U_{\vec{x}+(N_t -1) \cdot \hat{4},4} \right],
  \label{eq:B4O}
\end{equation}
with $\langle \cdots \rangle_{\rm c}$ the cumulants. 
Adopting the multi-point reweighting method, we first determine the transition line in the $(\beta,\lambda)$ plane 
as the minimum point of $B_4^\Omega$ for each $\lambda$, and then plot $B_4^\Omega$ on the transition line as continuous function of $\lambda$, as shown in Fig.~\ref{fig:B4}.
The right panel is an enlargement of the left panel around the crossing point.
Note that the precision is much improved over previous studies.

We find that, with the present high precision, $B_4^\Omega$ crosses at a point only when $LT\ge9$. 
To determine the crossing point, 
we fit the data by a FSS ansatz,
$
  B_4^\Omega(\lambda,LT) = b_4 + c ( \lambda-\lambda_{\rm c} ) (LT)^{1/\nu} ,
$
where $b_4$, $\lambda_{\rm c}$, $\nu$, $c$ are the fit parameters. 
The results of the fits using various ranges of $LT$ are shown by thick symbols in the right panel.
We find $\lambda_c = 0.00503(14)(2)$ with $b_4=1.630(24)(2)$ and $\nu=0.614(48)$, to be compared with the 3d Z(2) universality predictions $b_4\simeq1.604$ and $\nu\simeq0.630$.
This $\lambda_c$ corresponds to $\kappa_c=0.0602(4)$ for $N_f=2$.

\section{Scope and convergence of HPE \cite{wakabayashi}}
\label{sec:hpe}

In the study of critical point $\kappa_c$ for $N_t=4$ in the previous section, we confirm that the shift of $\kappa_c$ due to the NLO term is small ($\approx2.6$\%), suggesting that LO and NLO approximations are fairly accurate in this case \cite{Kiyohara:2021smr}.
On the other hand, $\kappa_c$ is known to become larger when we increase $N_t$ towards the continuum limit.
Therefore, before extending the study to larger $N_t$, we study the accuracy of the HPE.

\begin{figure}
  \centering
    \includegraphics[width=0.45\textwidth]{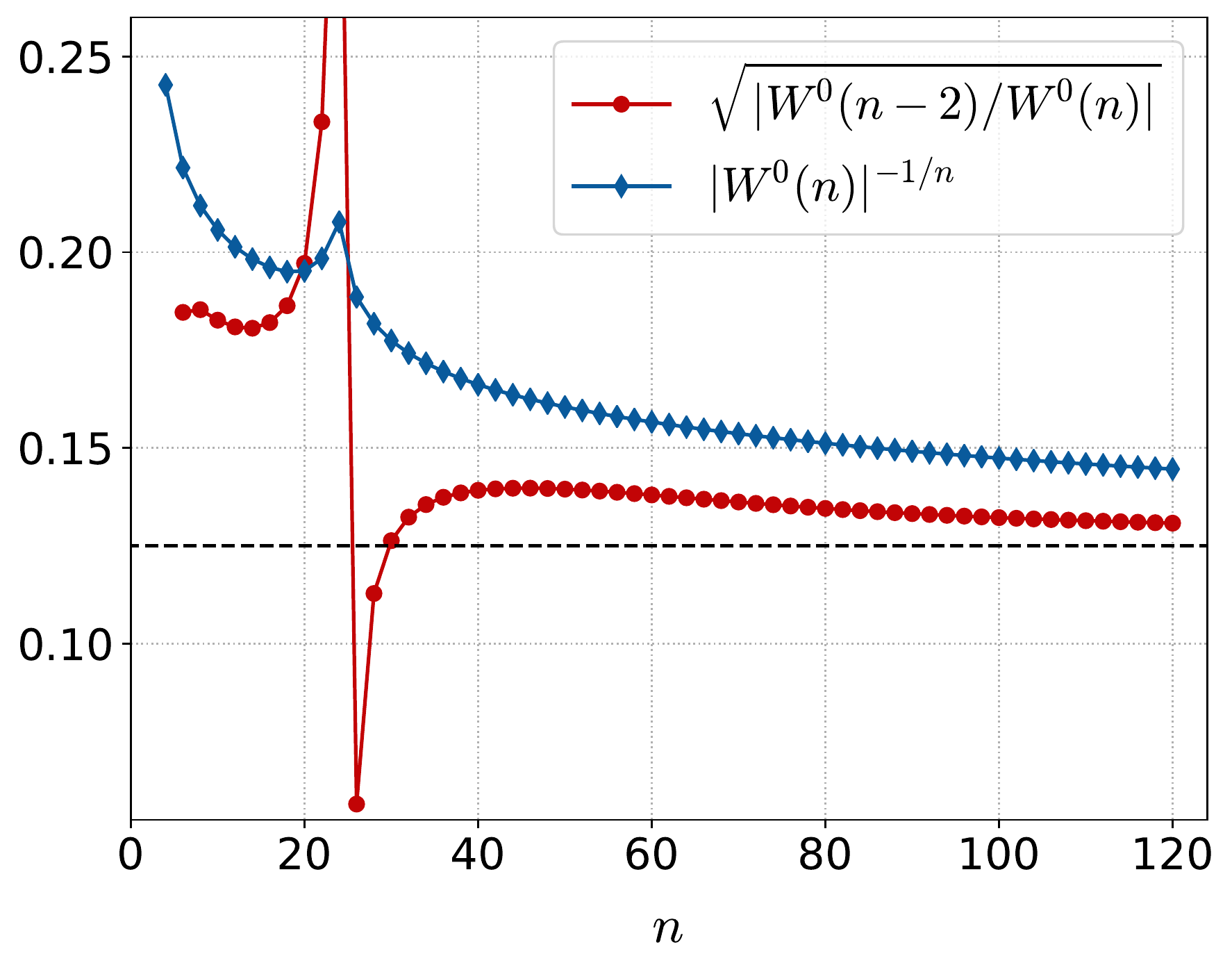}
    \hspace{1mm}
    \includegraphics[width=0.49\textwidth]{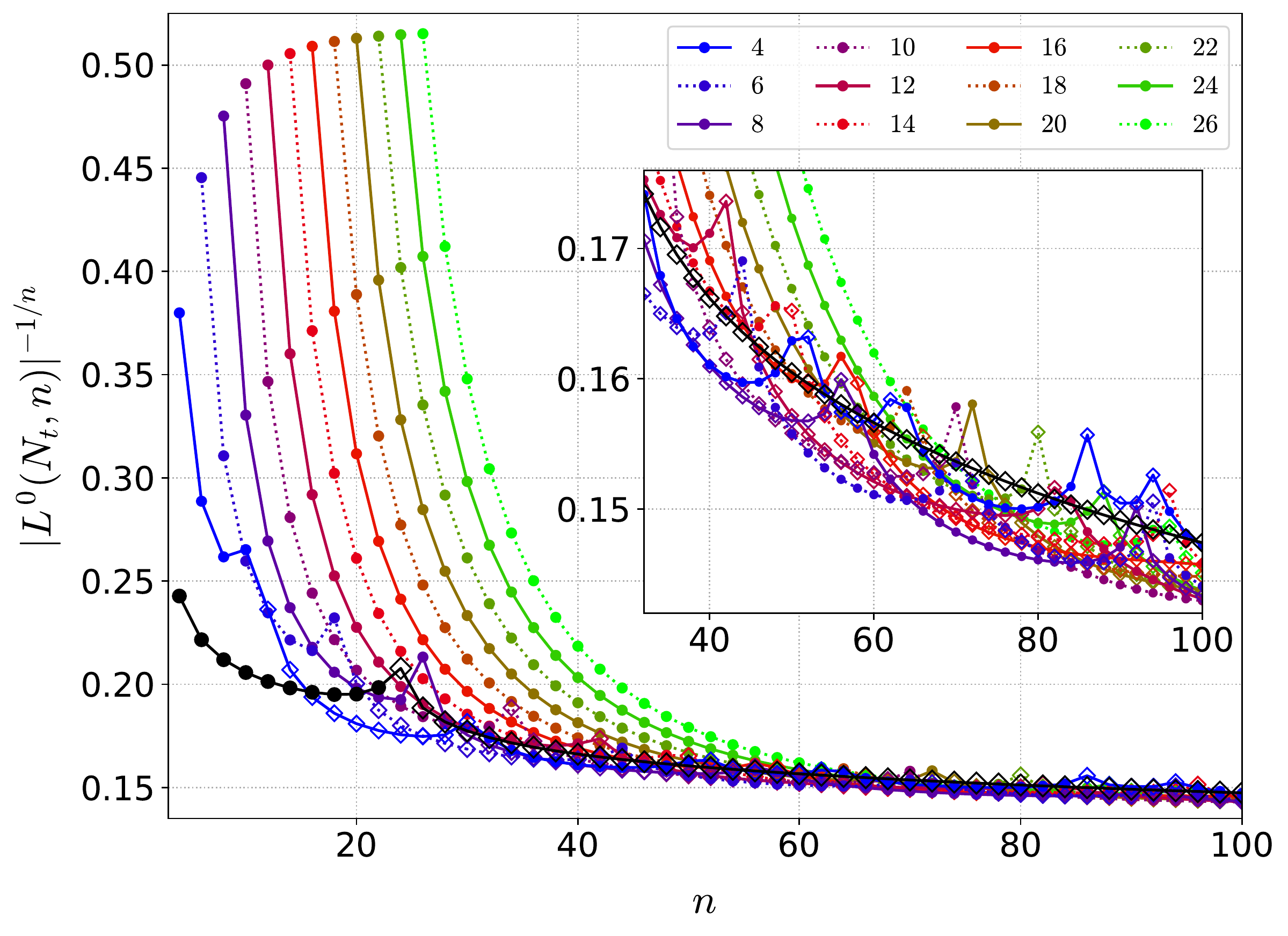}
    \vspace*{-0.5mm}
  \caption{
Convergence radius of the HPE for the worst convergent case \cite{wakabayashi}. 
The left panel shows the d'Alembert's radius $\kappa_{\rm dA}(W^0;n) = \sqrt{\left| W^0(n-2)/W^0(n) \right|}$ (red symbols) and the Cauchy-Hadamard's radius $\kappa_{\rm CH}(W^0;n) = \left| W^0(n) \right|^{-1/n} $ (blue symbols) for $\sum_n W^0(n)\,\kappa^n$. 
The horizontal dashed line represents the chiral limit $\kappa=1/8$ for free Wilson fermions.
The right panel shows the Cauchy-Hadamard's radius $\kappa_{\rm CH}(L^0;n) =  \left| L^0(N_t, n) \right|^{-1/n}$  for $\sum_n L^0(N_t,n)\,\kappa^n$
(colored symbols with $N_t$-dependent colors), together with $\kappa_{\rm CH}(W^0;n)$ (black symbols). 
The inset in the right panel is a close-up of the range $n=32$--100.}
\label{fig:convrad}
\end{figure}

We first note that, in the HPE of Eq.(\ref{eq:HPE}), the Wilson-type and the Polyakov-type loop operators take their maximum values when we set all link variables to unity.
Therefore, the worst convergent case of the HPE is provided by setting the link variables to unity. 
We denote $W(n)$ and $L_m(N_t,n)$ in this case as $W^0(n)$ and $L^0_m(N_t,n)$.
Because the dependence on the gauge configuration is removed in this case, we can calculate $W^0(n)$ and $L^0_m(N_t,n)$ analytically up to high orders \cite{wakabayashi}. 
In Fig.~\ref{fig:convrad}, we show results of the convergence radius for $\sum_n W^0(n)\,\kappa^n$ and $\sum_n L^0(N_t,n)\,\kappa^n$ up to $n=120$ and 100, respectively. 
We find that they approach $\kappa=1/8$ in the limit $n\to\infty$. 
We can show this also by an analytic calculation of $\det M(\kappa)$ itself in this case.
This result is understandable because we just have free Wilson quarks when we set the link variables to unity.

\begin{figure}
  \centering
    \includegraphics[width=0.324\textwidth]{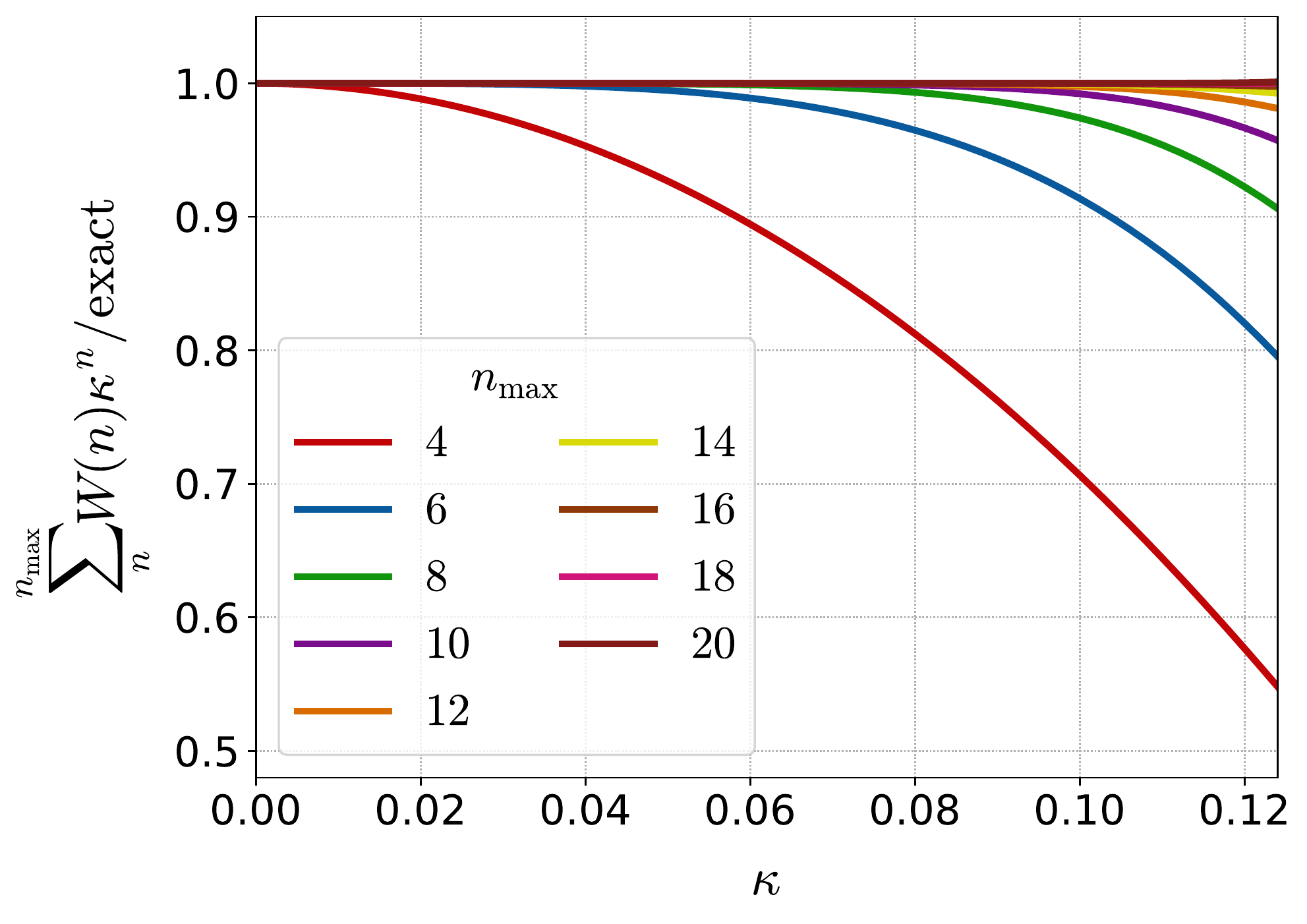}
    \includegraphics[width=0.324\textwidth]{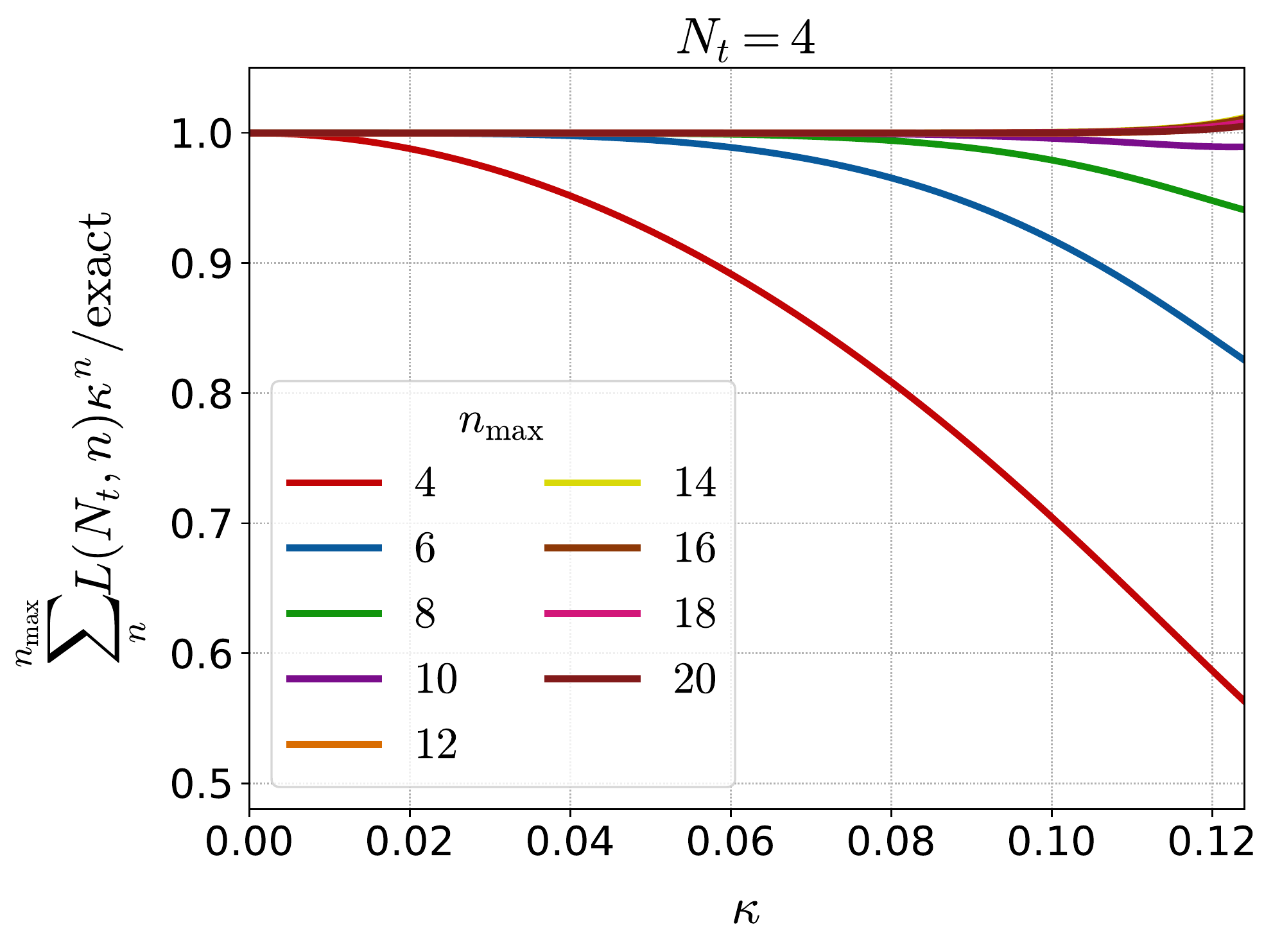}
    \includegraphics[width=0.324\textwidth]{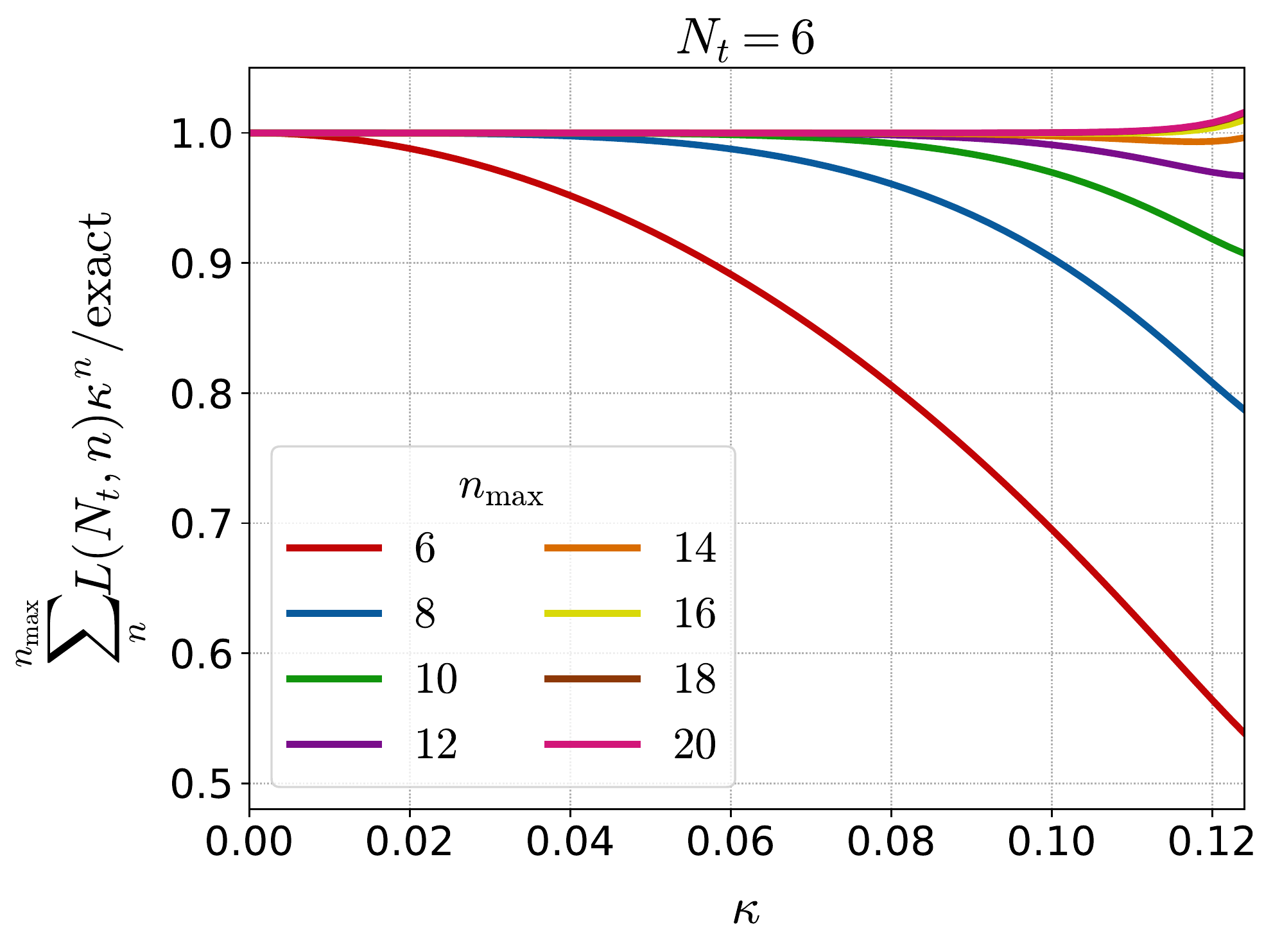}
    \vspace*{-0.5mm}
  \caption{
Relative deviation from the exact result of terms in the quark effective action when we truncate the HPE at finite order $n_{\rm max}$, in the worst convergent case.
Left panel is the results for the Wilson-type term $\sum_n^\infty W^0(n)\, \kappa^n$.
Central and right panels are for the Polyakov-type term $\sum_n^\infty L^0(N_t,n)\, \kappa^n$ at $N_t=4$ and 6, respectively.
Results at other $N_t$'s are similar.
These figures are reproduced using the data of \cite{wakabayashi}.
  }
\label{fig:scope}
\end{figure}

This means that the HPE is reliable up to the chiral limit when terms up to a sufficiently high order are incorporated.
Here, to which order we need to incorporate depends on the value of $\kappa$ we study and on the precision we require.
In the left panel of Fig.~\ref{fig:scope}, we show relative deviation from the exact result of Wilson-type contribution $\sum_n^\infty W^0(n) \kappa^n$ when we truncate $\sum_n$ at finite $n_{\rm max}$. 
The red curve is for the LO approximation ($n_{\rm max}=4$), and the blue curve the NLO approximation ($n_{\rm max}=6$). 
Results for the Polyakov-type contribution are similar, as shown in the central and right panels.
These deviations give the upper bounds of the truncation error in the real case of actual gauge configurations. 
From these figures, we see that, around the critical point $\kappa_c = 0.0602(4)$ for $N_t=4$, LO may have at worst $\approx10$\% systematic error due to the truncation, while NLO will help us in reducing the error smaller than 1--2\%. 
Around $\kappa_c = 0.0877(9)$ for $N_t=6$ \cite{Cuteri:2020yke}, NLO is required to ensure accuracy better than about 95\%, and around $\kappa_c = 0.1135(8)$ for $N_t=8$ \cite{Cuteri:2020yke}, NNLO is required to achieve a similar accuracy.

\section{Effective method to incorporate high orders of HPE \cite{wakabayashi}}
\label{sec:eff}

As the number of relevant loops increases exponentially with $n$, calculation of high order terms becomes quickly tedious with increasing $n$. 
In Ref.~\cite{wakabayashi}, we thus developed an effective method to incorporate expansion terms up to high orders 
by extending the idea of the effective NLO method of \cite{Ejiri:2019csa}.
Our basic observation is the strong correlation of Wilson-type and Polyakov-type loops among different $n$.

\begin{figure}
  \centering
    \includegraphics[width=0.35\textwidth]{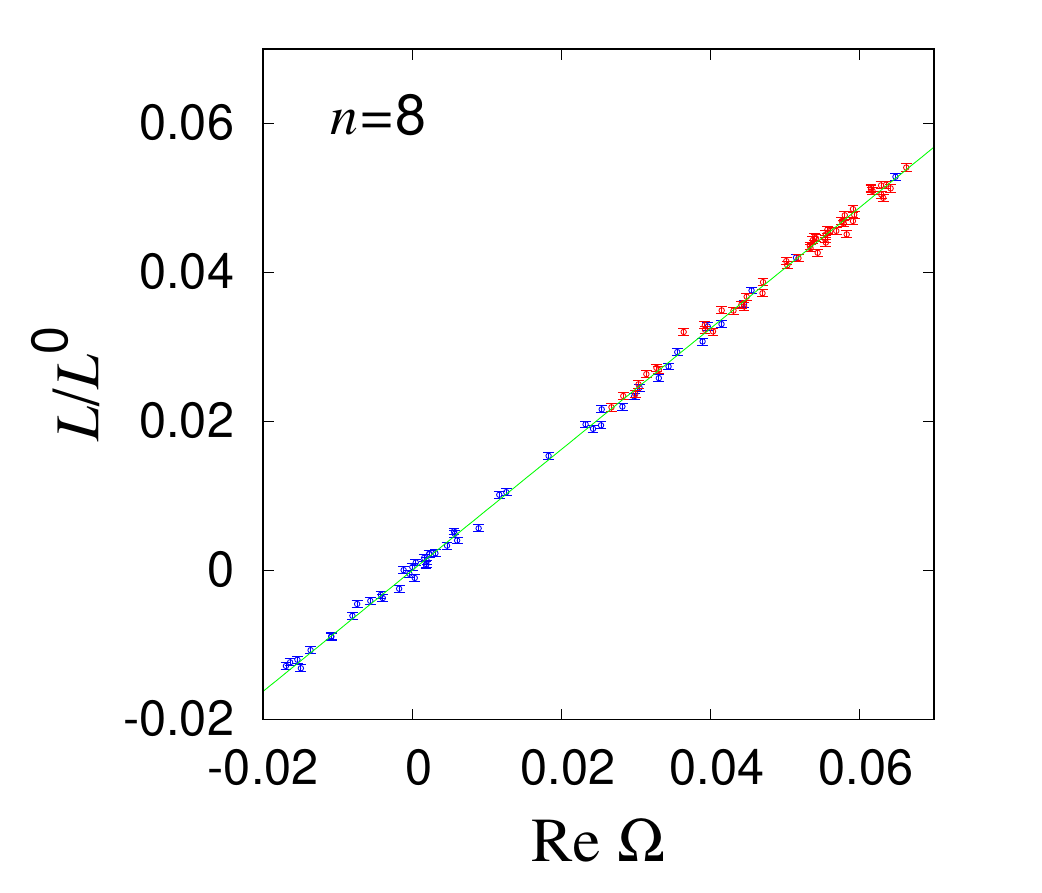}
    \hspace{-6mm}
    \includegraphics[width=0.35\textwidth]{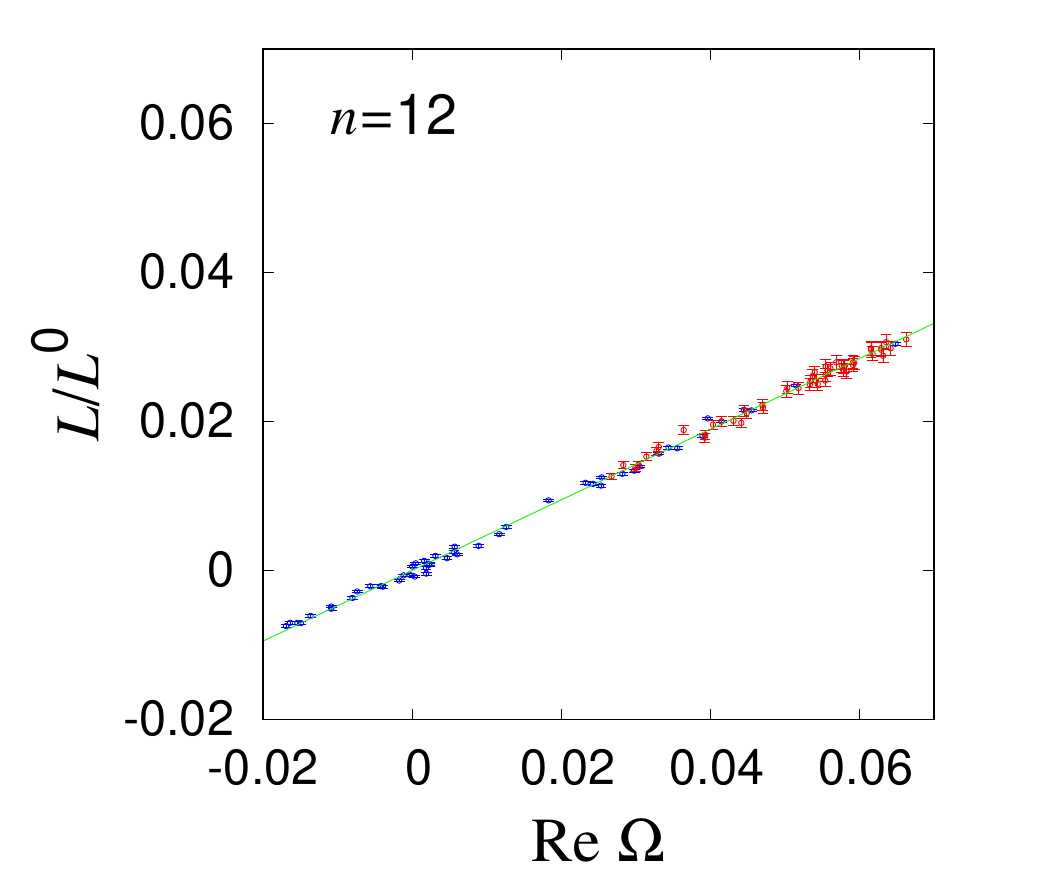}
    \hspace{-6mm}
    \includegraphics[width=0.35\textwidth]{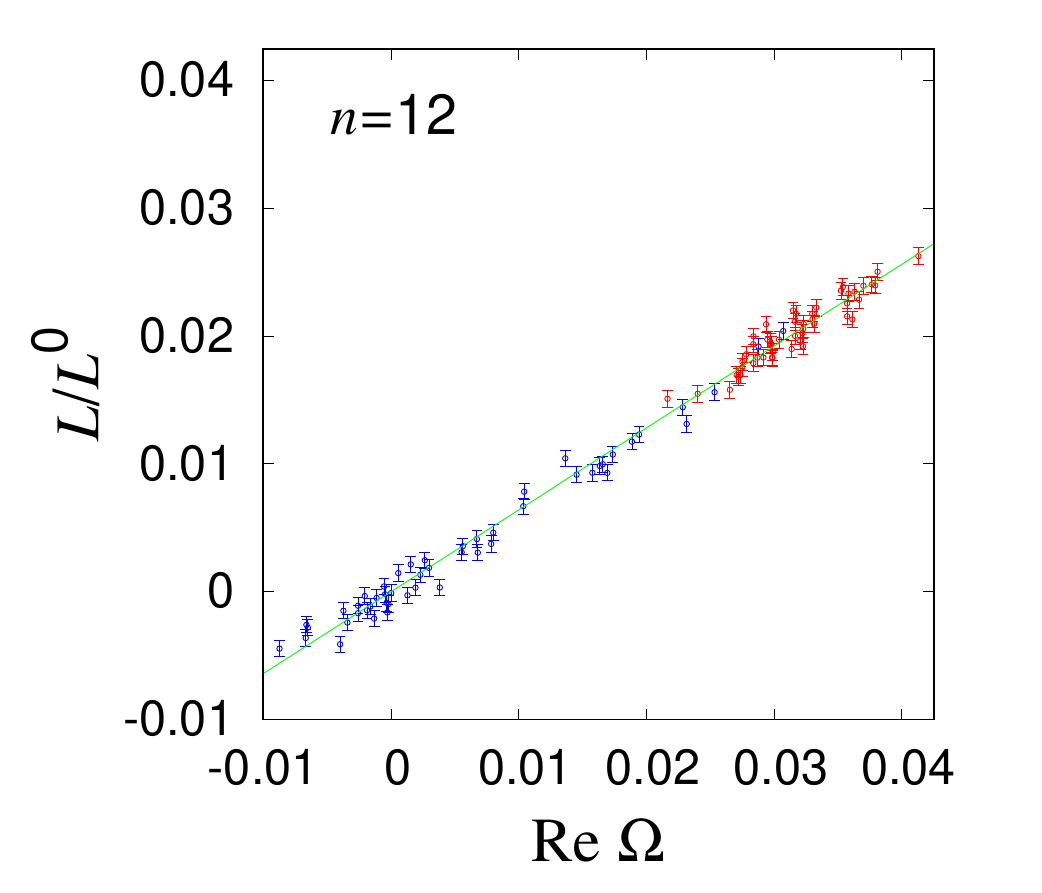}
    \hspace{-3mm}
    \vspace*{-0.5mm}
  \caption{
Double distributions of $L(N_t, n)/L^0(N_t,n)$ and ${\rm Re} \,\hat{\Omega}$ obtained on $32^3\times N_t$ lattices.
Left and central panels show the results for $n=8$ and 12 on the $N_t=6$ lattice, and the right panel for $n=12$ on the $N_t=8$ lattice.
The blue and red symbols are the data obtained at $\beta$ slightly below and above the transition point.
The green lines are the results of linear fits.
See~\cite{wakabayashi} for details.
  }
\label{fig:corr}
\end{figure}

We developed a method to separately evaluate $W(n)$ and $L_m(N_t,n)$ with general gauge configurations by combining the results of $D_n$ with various twisted boundary conditions, 
where $D_n$ itself can be evaluated by the noise method:
$
\textrm{Tr}\,[B^n] \approx \langle\!\langle \eta^\dag B^n \eta \rangle\!\rangle_\eta
$
with $\eta$ the noise vector.
Using configurations generated in the heavy quark limit on $32^3\times N_t$ lattices with $N_t=6$ and 8, we measure $W(n)$ and $L_m(N_t,n)$  around the transition point up to $n=20$.
Figure~\ref{fig:corr} shows typical double distributions of $L(N_t,,n)$ and ${\rm Re}\,\hat{\Omega}$. 
Strong linear correlation is visible up to $n=20$. 
We find that $W(n)$ also shows similar linear correlation with the plaquette, though the correlation is weaker than the Polyakov-type terms.

These suggest an approximation
\begin{equation}
L (N_t, n) \approx L^0 (N_t, n) \, c_n \, {\rm Re} \,\hat{\Omega} \, , 
\hspace{8mm}
W(n) \approx W^0 (n) \left(d_n \hat{P} + f_n\right),
\label{eq:approx}
\end{equation}
where the coefficients $c_n$, $d_n$ and $f_n$ are obtained by linear fits of the double distributions, 
and our results of them up to $n=20$ together with the values of $L^0(N_t,n)$ and $W^0(n)$ are given in~\cite{wakabayashi}.
Note that this approximation can be easily implemented in our LO configuration generation algorithm by just shifting the coupling parameters as 
\begin{equation}
\lambda \longrightarrow \lambda^* = N_f N_t \sum_{n=4}^{n_{\rm max}} L^0(N_t,n)\,c_n\kappa^n,
\hspace{5mm}
\beta \longrightarrow \beta^* = \beta + \frac{1}{6} N_f \sum_{n=4}^{n_{\rm max}} W^0(n) d_n \kappa^n.
\label{eq:approxLO}
\end{equation}
Using Eq.~(\ref{eq:approx}), we can also translate results of the critical point etc.\ with LO or NLO approximation to those effectively incorporating further higher order effects.

\section{Preliminary results at $N_t=6$}
\label{sec:nt6}

\begin{figure}
  \centering
    \includegraphics[width=0.47\textwidth]{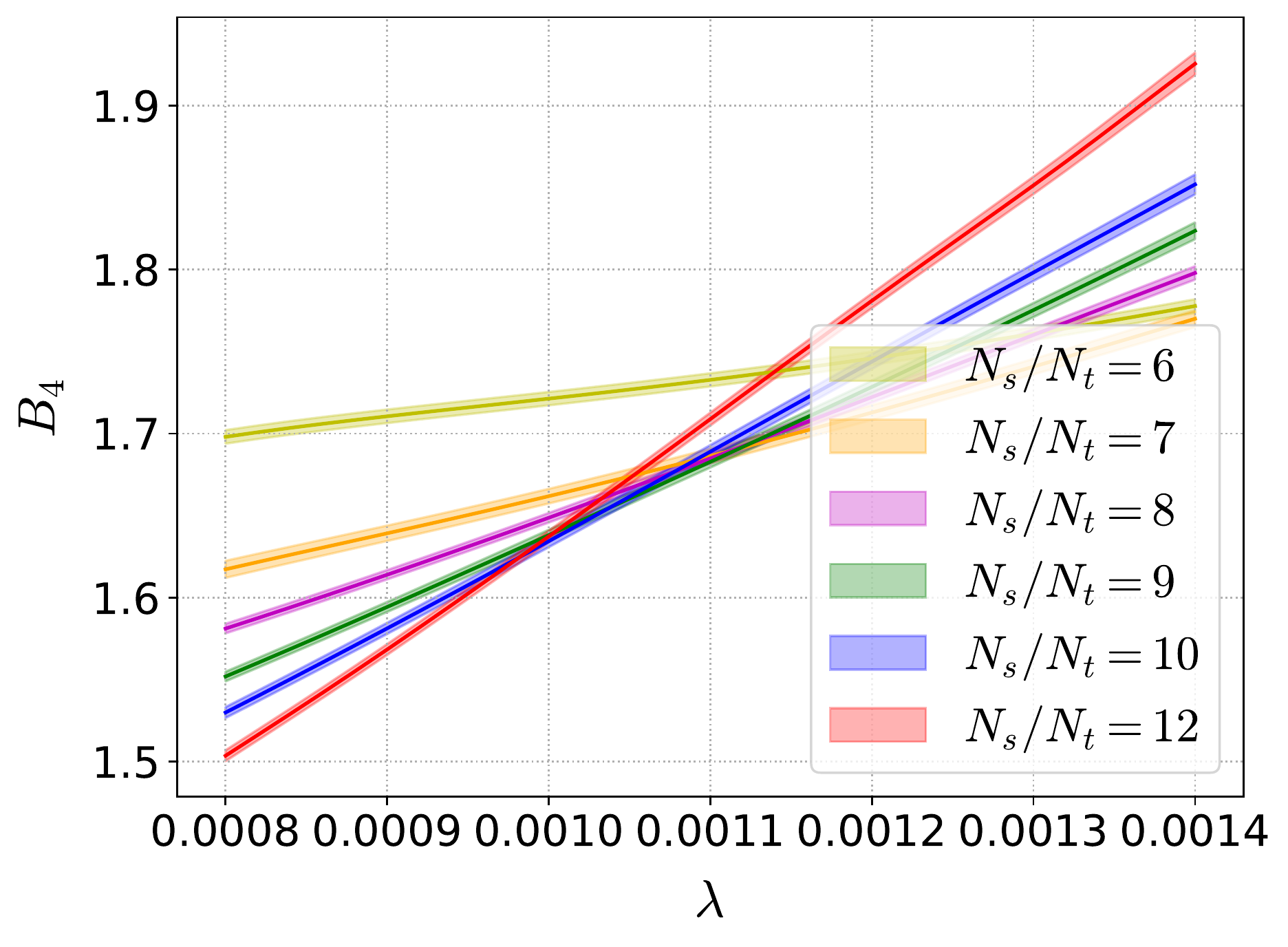}
    \hspace{1mm}
    \includegraphics[width=0.5\textwidth]{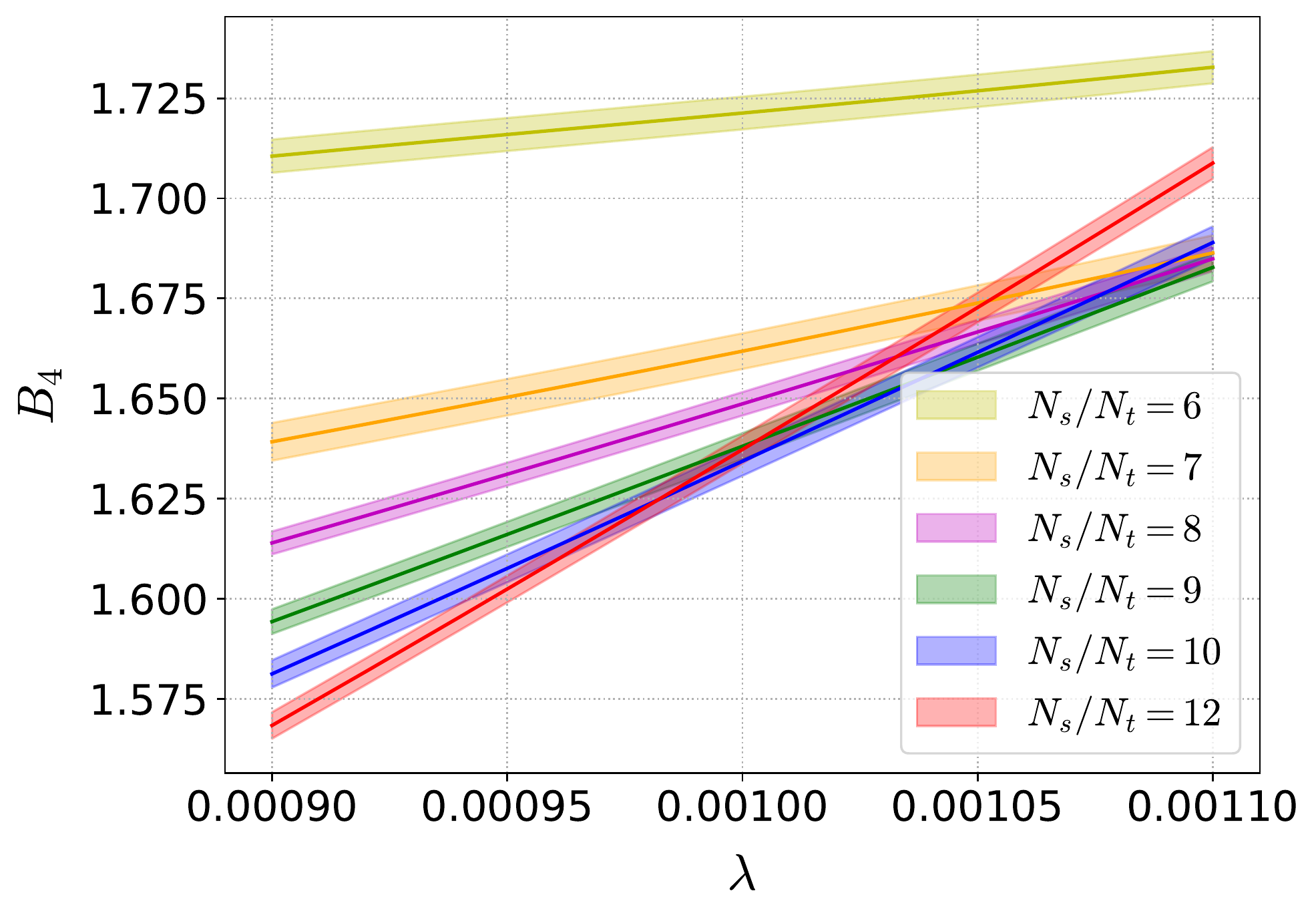}
    \vspace*{-0.5mm}
  \caption{
    Preliminary results of the Binder cumulant $B_4^\Omega$ to the NLO of HPE, obtained on $N_t=6$ lattices at six aspect ratios $LT=N_s/N_t=6$, $\cdots$, 12 .  The right panel is an enlargement of the left panel around the crossing point.
  }
\label{fig:B4-6}
\end{figure}

We are now extending our study to $N_t=6$ by generating configurations on $N_s^3\times6$ lattices with the aspect ratio $N_s/N_t=LT=6$, 7, 8, 9, 10 and 12, i.e., $N_s=36$, $\cdots$, 72. 
Our preliminary results of $B_4^\Omega$ using about $10^6$ configurations at each simulation point are shown in Fig.~\ref{fig:B4-6}. 
We see that, while the data at $N_s/N_t \le 8$ are clearly off the finite size scaling region, crossing points of the data at $N_s/N_t \ge 9$ are overlapping within the present errors. 
We also note that data at $N_s/N_t=6$--8 show larger violation of the finite size scaling than those on the $N_t=4$ lattices 
--- we may need larger $N_s/N_t$ than the case of $N_t=4$ to extract the large volume limit.
To identify the volume dependence more clearly, we are performing simulation at $N_s/N_t=15$. 

Assuming that the data at $N_s/N_t \ge 9$ are in the scaling region, we obtain $\lambda_c \sim 0.00098$--0.00104 in the NLO approximation, which correspond to $\kappa_c \sim 0.093$-0.094 for $N_f=2$.
Incorporating the effect of higher order terms up to $n=20$ using Eq.~(\ref{eq:approx}), we then obtain $\kappa_c \sim 0.091$, to be compared with $\kappa_c=0.0877(9)$ by a full QCD simulation on lattices with $N_s/N_t=4$--7 \cite{Cuteri:2020yke}.

\section{Summary}
\label{sec:summary}

We studied the critical point of finite-temperature QCD in the heavy quark region on lattices with large spatial volumes. 
To carry out high-statistic simulations with large spatial volumes, we adopt the hopping parameter expansion (HPE). 
We also adopt the multi-point reweighting method to continuously vary the coupling parameters in the scaling study around the critical point.
We include the leading order (LO) terms of the HPE in the configuration generation to solve the overlap problem of the reweighting method, and take the next-leading order (NLO) terms in the measurements by reweighting. 
From a study of the Binder cumulant on $N_t=4$ lattices with large spatial volumes, we find that the finite size scaling around the critical point is realized only when the spatial size is quite large -- in terms of the aspect ratio, $LT=N_s/N_t \ge 9$ is required to our precision, where $T$ is the transition temperature in this case. 
Scaling fit using data on large lattices enabled us to determine the critical point in the thermodynamic limit with high precisions. 
We found $\kappa_c=0.0602(4)$ for $N_f=2$ QCD with the Binder cumulant and the critical exponent consistent with the expected Z(2) universality class.

To confirm the validity of the HPE, we then studied the convergence radius in the worst convergent case of the HPE.
We found that the HPE is reliable up to the chiral limit when terms up to a sufficiently high order are incorporated.
From a study of the truncation error of the HPE, we confirmed that, around the critical point, the LO (NLO) approximation of the HPE is fairly accurate for $N_t=4$ (6), while we need to incorporate higher order effects for larger $N_t$. 
To extend the study to large $N_t$ lattices, we thus developed a method to take higher-order terms up to a sufficiently high order. 

Finally, we reported on the status of our current project to determine $\kappa_c$ for $N_t=6$.
We found that the violation of the finite size scaling is larger than the case of $N_t=4$. 
Our preliminary result for the critical point in $N_f=2$ QCD, estimated by the crossing point of the Binder cumulant on $N_s/N_t=9$--12 lattices, is $\kappa_c \sim 0.093$-0.094 in the NLO approximation. 
Using the effective method we developed, this corresponds to $\kappa_c \sim 0.091$ when we incorporate the effect of higher order terms of the HPE up to the 20th order.
To identify the volume dependence more clearly, we are performing simulation at $N_s/N_t=15$. 

It is easy to generalize our effective method to QCD at non-zero densities.
Study towards this direction is also in progress, besides investigations of various thermodynamic quantities using the SF$t$X method based on the gradient flow \cite{sftx,sftx2}.

\section*{Acknowledgments}

We thank the members of the WHOT-QCD Collaboration for valuable discussions.
This work was in part supported by JSPS KAKENHI Grant Numbers JP22K03595, JP22K03619, JP21K03550, JP20H01903,
JP19H05146, JP19H05598, and JP19K03819. 
This research used computational resources provided by the Interdisciplinary Computational Science Program of Center for Computational Sciences, University of Tsukuba,
and by the JHPCN through HPCI System Research Projects (Project ID:hp200013, hp200089, hp210012, hp210039, hp220020, hp220024) and JHPCN projects (jh200010, jh200049), OCTOPUS and SQUID at Cybermedia Center, Osaka University, ITO at Research Institute for Information Technology, Kyushu University, and Grand Chariot at Information Initiative Center, Hokkaido University.


\end{document}